\documentclass[10pt,twoside,twocolumn,english,aps,manuscript,aps,manuscript,preprint,show pacs,superscriptaddress]{revtex4}
\usepackage[T1]{fontenc}
\usepackage[latin9]{inputenc}
\usepackage{color}
\usepackage{babel}
\usepackage{amstext}
\usepackage{graphicx}
\usepackage[unicode=true,pdfusetitle,
 bookmarks=true,bookmarksnumbered=false,bookmarksopen=false,
 breaklinks=false,pdfborder={0 0 1},backref=false,colorlinks=false]
 {hyperref}

\makeatletter

\providecommand{\tabularnewline}{\\}

\@ifundefined{textcolor}{}
{%
 \definecolor{BLACK}{gray}{0}
 \definecolor{WHITE}{gray}{1}
 \definecolor{RED}{rgb}{1,0,0}
 \definecolor{GREEN}{rgb}{0,1,0}
 \definecolor{BLUE}{rgb}{0,0,1}
 \definecolor{CYAN}{cmyk}{1,0,0,0}
 \definecolor{MAGENTA}{cmyk}{0,1,0,0}
 \definecolor{YELLOW}{cmyk}{0,0,1,0}
}

\makeatother

\begin{document}

\title{Unconventional magnetism in the spin-orbit driven Mott insulators
Ba$_{3}$\textit{M}Ir$_{2}$O$_{9}$ (\textit{M}=Sc,Y)}

\author{Tusharkanti Dey }

\altaffiliation[Present address: ]{Leibniz-Institute for Solid State and Materials Research, IFW Dresden, 01171 Dresden, Germany}

\selectlanguage{english}%

\affiliation{Department of Physics, Indian Institute of Technology Bombay, Powai,
Mumbai 400076, India}

\author{R. Kumar}

\affiliation{Department of Physics, Indian Institute of Technology Bombay, Powai,
Mumbai 400076, India}

\author{A.V. Mahajan}

\email{mahajan@phy.iitb.ac.in}

\selectlanguage{english}%

\affiliation{Department of Physics, Indian Institute of Technology Bombay, Powai,
Mumbai 400076, India}

\author{S. D. Kaushik}

\affiliation{UGC-DAE Consortium for Scientific Research, Mumbai Centre, R-5 Shed,
Bhabha Atomic Research Centre, Mumbai 400085, India}

\author{V. Siruguri}

\affiliation{UGC-DAE Consortium for Scientific Research, Mumbai Centre, R-5 Shed,
Bhabha Atomic Research Centre, Mumbai 400085, India}
\begin{abstract}
We have carried out detailed bulk and local probe studies on the hexagonal
oxides Ba$_{3}$\textit{M}Ir$_{2}$O$_{9}$ (\textit{M}=Sc,Y) where
Ir is expected to have a fractional oxidation state of $+4.5$. In
the structure, Ir-Ir dimers are arranged in an edge shared triangular
network parallel to the $ab$ plane. Whereas only weak anomalies are
evident in the susceptibility data, clearer anomalies are present
in the heat capacity data; around $10$K for Ba$_{3}$ScIr$_{2}$O$_{9}$
and at $4$K for Ba$_{3}$YIr$_{2}$O$_{9}$. Our $^{45}$Sc nuclear
magnetic resonance (NMR) lineshape (first order quadrupole split)
is symmetric at room temperature but becomes progressively asymmetric
with decreasing temperatures indicating the presence of developing
inequivalent Sc environments. This is suggestive of distortions in
the structure which could arise from progressive tilt/rotation of
the IrO$_{6}$ octahedra with a decrease in temperature $T$. The
$^{45}$Sc NMR spectral weight shifts near the reference frequency
with decreasing $T$ indicating the development of magnetic singlet
regions. Around 10K, a significant change in the spectrum takes place
with a large intensity appearing near the reference frequency but
with the spectrum remaining multi-peak. It appears from our $^{45}$Sc
NMR data that in Ba$_{3}$ScIr$_{2}$O$_{9}$ significant disorder
is still present below $10$K. In the case of Ba$_{3}$YIr$_{2}$O$_{9}$,
the $^{89}$Y NMR spectral lines are asymmetric at high temperatures
but become nearly symmetric (single magnetic environment) below $T\sim70$K.
Our $^{89}$Y spectra and $T_{1}$ measurements confirm the onset
of long range ordering (LRO) from a bulk of the sample at $4$K in
this compound. Our results suggest that Ba$_{3}$YIr$_{2}$O$_{9}$
might be structurally distorted at room temperature (via, for example,
tilt/rotations of the IrO$_{6}$ octahedra) but becomes progressively
a regular triangular lattice with decreasing $T$. The effective magnetic
moments and magnetic entropy changes are strongly reduced in Ba$_{3}$YIr$_{2}$O$_{9}$
as compared to those expected for a $S=1/2$ system. Similar effects
have been found in other iridates which naturally have strong spin-orbit
coupling (SOC). 
\end{abstract}


\maketitle

\section{introduction}

Recently, iridium ($5d$) based oxides have been the attraction of
materials researchers due to their interesting quantum states \cite{G.Cao book}.
In the case of $3d$ based oxides, a large onsite Coulomb interaction
(\emph{U}) and crystal field splitting are the driving interactions
and a small spin-orbit coupling (SOC) is generally considered as a
perturbation. The large \emph{U} drives them to a Mott insulating
state. In the case of iridates, \emph{U }is is expected to be much
smaller because of the extended nature of the $5d$-orbitals and one
might expect the iridates to be metallic rather than magnetic. In
contradiction to this naive expectation, many iridates like Sr$_{2}$IrO$_{4}$
\cite{Kim-PRL-101-2008,Kim-Science-2009}, (Na/Li)$_{2}$IrO$_{3}$
\cite{Cao-Na2IrO3-PRB(R),Gretarsson-PRL-2013-A2IrO3,Manni-A2IrO3-arxiv},
Na$_{4}$Ir$_{3}$O$_{8}$ \cite{Okamoto-PRL-99-2007-Na4Ir3O8,Singh-PRB-Na4Ir3O8},
and Ba$_{3}$IrTi$_{2}$O$_{9}$ \cite{Dey-PRB-2012-Ba3IrTi2O9} are
found to be insulators showing exotic magnetic properties. In these
iridates, SOC is believed to be comparable to \emph{U}. The interplay
of SOC and \emph{U} makes these materials Mott insulators through
a different mechanism and are often called spin-orbit driven Mott
insulators. Kim \textit{et al.} \cite{Kim-PRL-101-2008,Kim-Science-2009}
first reported such a spin-orbit driven Mott insulating state in Sr$_{2}$IrO$_{4}$.
Here a large SOC splits the $t_{2g}$ orbitals of Ir$^{4+}$($5d^{5}$)
into a half filled $J_{eff}=1/2$ doublet and a completely filled
$J_{eff}=3/2$ quadruplet. Further, even a small \emph{U} can split
the narrow $J_{eff}=1/2$ band into lower and upper Hubbard bands
and the material becomes a spin-orbit driven Mott insulator. Rapid
theoretical \cite{Yang-PRB-2011-WehylMetal,Khaliullin-PRL-d4,Chaloupka-PRL-2010,Wang-PRL-2011,Chen-PRB-82-2010}
and experimental advances are being made in this area. All the materials
mentioned above have Ir$^{4+}$($5d^{5}$) oxidation state. Materials
with iridium oxidation state other than $4+$ are not explored much
but could be interesting.

We have been searching for new candidates to study this spin-orbit
driven physics further. We found the triple perovskite Ba$_{3}$\emph{M}Ir$_{2}$O$_{9}$
series to be rather interesting because a wide range of elements with
a variety of oxidation states can be fitted as \emph{M}. Several reports
are available where \emph{M} is an alkaline metal, a transition metal
or a rare earth material. The nominal oxidation state of Ir can be
$5.5+$ (for \emph{M}=Li$^{+}$, Na$^{+}$ \cite{Kim-JSSC-177-2004-Ba3LiIr2O9}),
$5+$ (for \emph{M}=Zn$^{2+}$, Mg$^{2+}$, Ca$^{2+}$, Cd$^{2+}$
\cite{Sakamoto-JSSC-179-2006}), $4.5+$ (for \emph{M}=Y$^{3+}$,
Sc$^{3+}$, In$^{3+}$, Lu$^{3+}$ \cite{Sakamoto-JSSC-179-2006,Doi-JPCM-16-2004})
and $4+$ (for \emph{M}=Ti$^{4+}$, Zr$^{4+}$ \cite{Sakamoto-JSSC-179-2006}).
In all these compounds, Ir-Ir structural dimers are formed parallel
to the crystallographic $c$-axis and these dimers are connected with
each other to form an edge shared triangular lattice in the $ab$-plane.
Novel properties are expected from this structural arrangement in
addition to those driven by SOC. 

The materials with a fractional oxidation state of Ir are even more
interesting because the fractional oxidation state of Ir (accompanied
by a unique crystallographic site) can lead them to a metallic state.
Sakamoto \textit{et al.} \cite{Sakamoto-JSSC-179-2006} reported results
of magnetic susceptibility and heat capacity measurements on Ba$_{3}$Sc$^{3+}$Ir$_{2}^{4.5+}$O$_{9}$.
An anomaly at $10$K was seen in the heat capacity data (although
it was not as sharp as normally seen for long-range ordering). The
temperature derivative of the susceptibility also showed an anomaly
at the same temperature. They suggested that the anomalies arose from
the antiferromagnetic (AF) ordering of the spins. A Curie-Weiss (CW)
fitting of the susceptibility data yielded a large AF Weiss temperature
$\theta_{\mathrm{CW}}=-570$K and an effective magnetic moment $\mu_{eff}=1.27\mu_{\mathrm{B}}$/Ir.
Such a large AF $\theta_{\mathrm{CW}}$ and a much reduced magnetic
ordering temperature indicates presence of frustration in the system.
On the other hand, the Y-analog, Ba$_{3}$Y$^{3+}$Ir$_{2}^{4.5+}$O$_{9}$
\cite{Doi-JPCM-16-2004} showed a sharper anomaly at $4$K in the
heat capacity data although no such feature was seen in the susceptibility
data. A CW fitting of the susceptibility data in this case was not
found to be satisfactory. A low moment found in both the cases was
explained by formation of antiferromagnetic dimers \cite{Sakamoto-JSSC-179-2006,Doi-JPCM-16-2004}.
However a characteristic broad maximum followed by an exponential
fall of susceptibility with decreasing temperature was not seen in
either of the cases. These facts inspired us to take up the study
of these two materials (\emph{i.e., }\textit{\emph{with }}\emph{M}=Y,
Sc) in detail. When reacted under high pressure, Ba$_{3}$YIr$_{2}$O$_{9}$
becomes cubic and the long-range ordering (LRO) disappears. We have
earlier reported our results on the high pressure phase of Ba$_{3}$YIr$_{2}$O$_{9}$
and compared our results with the ambient pressure phase \cite{Dey-PRB-2013-Ba3YIr}.
We have suggested that the high pressure phase is a spin-orbit driven
gapless spin-liquid.

Nuclear magnetic resonance (NMR) is a very useful technique and widely
used in strongly correlated electron systems to understand various
physics issues like spin gap \cite{Bobroff-PRL-2009-BiCu2PO6,Iwase-JPSJ-Cav2O5,Julien-PRL-spingap,Lue-PRB-2009-BiCu2VO6},
magnetic ordering \cite{Libu-PRB-LiCrO2,Gippius-PRB-2004-LiCu2O2,Ohsugi-PRL-1999-ordering-spinladder,Nath-PRB-2005},
superconductivity \cite{Bobroff-PRL-NMR-cuprate,Nakai-PRL-2010-BaFe2As-P,Ishida-Review-FeAs}
etc. However in this newly developing field of SOC driven materials,
to the best of our knowledge, very few \cite{Dey-PRB-2013-Ba3YIr,Takagi-PPT,Aharen-PRB-81-2010-Ba2YMoO6,Aharen-PRB-2009-Ba2YRuO6}
NMR results have been reported. As iridium is a good absorber of neutrons,
iridates are not easily probed by neutron diffraction/scattering.
In these Ba$_{3}$\emph{M}Ir$_{2}$O$_{9}$ (\emph{M}=Sc, Y) materials,
NMR could be a very useful local probe to throw light on their magnetism.
In fact in the present case, while we found magnetic ordering from
$^{89}$Y NMR measurements in Ba$_{3}$YIr$_{2}$O$_{9}$, ordering
could not be detected from neutron diffraction experiments. Also,
as discussed earlier, an exponential fall of susceptibility (characteristic
of antiferromagnetic dimer systems) was not seen in the case of these
materials. Sometimes in bulk measurements the intrinsic susceptibility
gets dominated by extrinsic effects. Using a local probe like NMR
enables one to track the intrinsic susceptibility via the NMR shift
measurement \cite{Quilliam-PRL-2012-Ba3CuSb2O9,Imai-PRL-2008}. Here
in this paper, we report the results of our magnetization, heat capacity,
and NMR measurements ($^{45}$Sc and $^{89}$Y) on ambient pressure
phases of Ba$_{3}$\emph{M}Ir$_{2}$O$_{9}$ (\emph{M}=Sc, Y). In
addition to the bulk measurements, our NMR measurements confirm LRO
at $4$K for Ba$_{3}$YIr$_{2}$O$_{9}$. While the $^{89}$Y NMR
spectrum at high temperature seems typical of a powder pattern with
axially symmetric shift anisotropy, the asymmetry decreases with decreasing
temperature. One could also speculate that there is distribution of
Y environments at room temperature which evolves into a single environment
with a decrease in temperature, possibly changing from a distorted
triangular lattice at high temperature to a near-perfect triangular
lattice at lower temperatures with eventual ordering near 4K. On the
other hand, in Ba$_{3}$ScIr$_{2}$O$_{9}$, no anomaly is seen in
the magnetization data while only a miniscule difference is present
in the zero field cooled (ZFC) and field cooled (FC) susceptibilities
in a $100$Oe field below about $8$K. A weak and broad heat capacity
anomaly is seen around this temperature. Our $^{45}$Sc NMR measurements
in Ba$_{3}$ScIr$_{2}$O$_{9}$ indicate that at room temperature,
there is a single local environment for Sc suggesting a near-perfect
triangular lattice. With decreasing temperature, the $^{45}$Sc NMR
line becomes asymmetric with increasing weight around the reference
frequency. This could arise from progressive distortions in the structure
in some regions (for instance, via tilt/rotation of the IrO$_{6}$
octahedra) leading to deviations from magnetic triangular lattice.
The progressive increase of the spectral weight near the reference
frequency is suggestive of a crossover to a singlet state. Finally,
below about 10K, there is a significant change in the lineshape with
a large peak near the reference frequency. In accord with this, the
$^{45}$Sc NMR spin-lattice relaxation rate $1/T_{1}$ corresponding
to the zero-shifted line becomes small at low temperature, indicating
the absence of magnetic fluctuations. The spectrum, however, remains
multi-peak and suggests the presence of significant disorder.

\begin{figure}
\centering{}\includegraphics[scale=0.6]{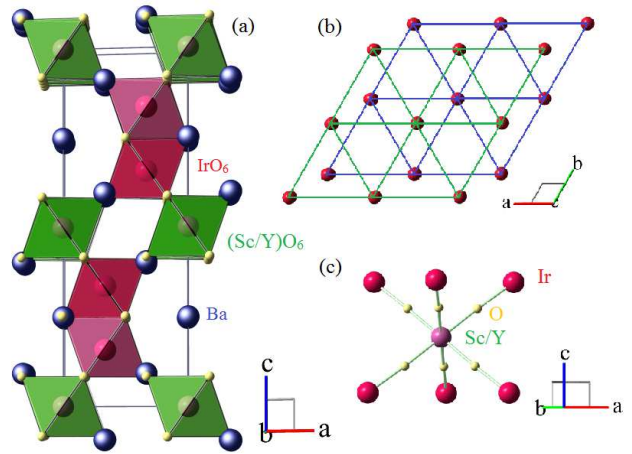}\caption{\label{fig:Ba3ScIr2O9-Unitcell}(a) The unit cell of the Ba$_{3}$(Y/Sc)Ir$_{2}$O$_{9}$
system. (b) The edge shared triangular network in the $ab$ plane
formed by the Ir-Ir dimers is shown. (c) The Sc/Y environment is shown.}
\end{figure}

\section{experimental details}

Polycrystalline samples of Ba$_{3}$\emph{M}Ir$_{2}$O$_{9}$ (\emph{M}=Sc,Y)
were prepared by solid state reaction method as detailed in Refs.
\cite{Sakamoto-JSSC-179-2006} and \cite{Doi-JPCM-16-2004}. We have
used high purity BaCO$_{3}$, Sc$_{2}$O$_{3}$/Y$_{2}$O$_{3}$ and
Ir metal powder as starting materials. Sc$_{2}$O$_{3}$ and Y$_{2}$O$_{3}$
powders were pre-dried at $1000^{0}$C overnight. Stoichiometric amount
of the powders were mixed thoroughly, pressed into pellets and calcined
at $900^{0}$C for $12$h. After calcination, the pellet was crushed
into powder and again pelletized and fired at $1300^{0}$C for $100$h
with intermediate grindings.

Powder x-ray diffraction (xrd) with Cu $K_{\alpha}$ radiation ($\lambda=1.54182\textrm{\AA}$)
and neutron diffraction ($\lambda=1.48\textrm{\AA}$) measurements
(for Ba$_{3}$YIr$_{2}$O$_{9}$) were carried out at room temperature
for structural characterization. The magnetization $M$ measurements
were done in the temperature (\emph{T}) range $2-400$K and field
(\emph{H}) range $0-70$kOe using a Quantum Design SQUID VSM. Heat
capacity measurements were performed using the heat capacity attachment
of a Quantum Design PPMS. The $^{45}$Sc and $^{89}$Y NMR measurements
were done in a fixed magnetic field of $93.954$kOe obtained inside
a room-temperature bore Varian superconducting magnet. We have used
a Tecmag pulse spectrometer for our measurements. For variable temperature
measurements, an Oxford continuous flow cryostat was used with liquid
nitrogen in the temperature range $80-300$K and liquid helium in
the temperature range $4-80$K. The $^{45}$Sc and $^{89}$Y nuclear
parameters are shown in Table \ref{tab:Nucleus}. Spectral lineshape
was obtained by Fourier transform of the spin echo signal resulting
from a $\pi/2-\tau-\pi$ pulse sequence. Spin-lattice relaxation time
was measured by the saturation recovery method. Spin-spin relaxation
measurements were carried out using a standard $\pi/2-t-\pi$ pulse
sequence. Spin-spin relaxation time ($T_{2}$) was obtained by fitting
the time dependence of spin-echo intensity $m(t)$ with 
\begin{equation}
m(t)=m(0)\, exp(-2t/T_{2})\label{eq:T2fittingFormula}
\end{equation}
Since the bulk data for the title compounds have been reported earlier,
we enumerate below our main conclusions concerning the xrd, magnetization
and heat capacity measurements. To avoid repetition, only relevant
figures are shown where the analysis is standard.

\begin{figure}
\raggedright{}\includegraphics[scale=0.6]{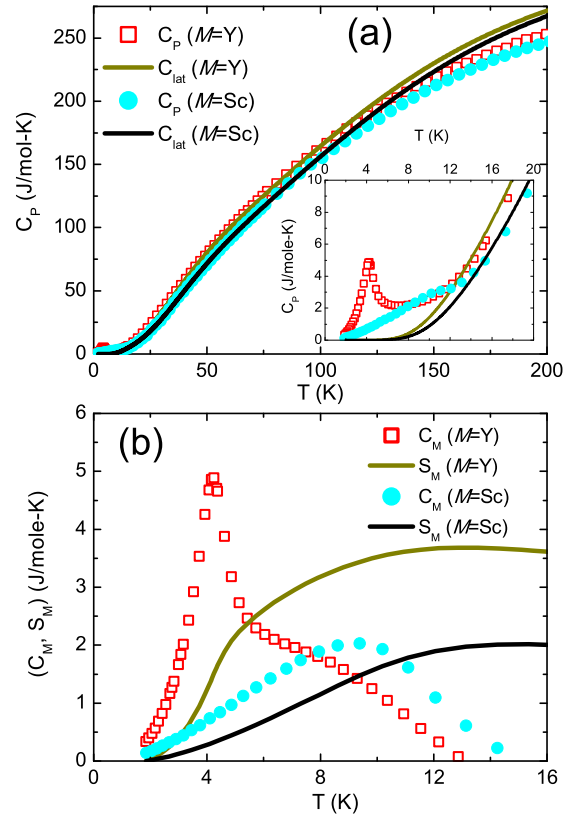}\caption{\label{fig:CombinedHC} (a) The heat capacities ($C_{\mathrm{P}}$)
of Ba$_{3}$\emph{M}Ir$_{2}$O$_{9}$ (\emph{M}=Sc,Y) samples measured
in zero field and their fitting with the Debye model (solid lines)
are shown. Inset: The low-temperature part of the $C_{\mathrm{P}}$
data are shown along with the extrapolated Debye fit curve. (b) The
magnetic heat capacity $C_{\mathrm{M}}$ (per mole formula unit) obtained
by subtracting the lattice part from $C_{\mathrm{P}}$ is shown. Typical
error bars are indicated for a few points. The corresponding entropy
changes ($\triangle S_{\mathrm{M}}$) for both the samples are also
shown.}
\end{figure}

\begin{enumerate}
\item We were able to refine our xrd data for both the compounds corresponding
to the space group P6$_{3}$/mmc. From our refinement, we found a
small ($\sim6\%$) site disorder between Sc$^{3+}$ ions at the $2a$
site and Ir$^{4.5+}$ ions at the $4f$ site for Ba$_{3}$ScIr$_{2}$O$_{9}$
which is similar to the site disorder ($5.5\%$) found by Sakamoto
\textit{et al.} \cite{Sakamoto-JSSC-179-2006}. Figs. \ref{fig:Ba3ScIr2O9-Unitcell}(a)
and (b) illustrate the main feature of the structure, namely, Ir-Ir
dimers forming edge shared triangles. The Y/Sc environment is demonstrated
in Fig. \ref{fig:Ba3ScIr2O9-Unitcell}(c).
\item For both the compounds, the susceptibility has a Curie-Weiss (CW)
nature, $\chi=\chi_{0}+C/(T-\theta_{\mathrm{CW}})$. Also, while the
Sc-compound shows no anomaly in $\chi(T)$, a weak anomaly at about
$4$K was observed in the Y-compound (see Fig. \ref{fig:Ba3YIr2O9-Shift}).
Note however that for the Sc-compound the susceptibility has very
little $T$-dependence down to nearly $50$K (see Fig. \ref{fig:Ba3ScIr2O9-Kchi})
and hence the $\theta_{\mathrm{CW}}$ and $C$ values obtained for
this compound should be taken with caution. In any case, the effective
moments ($\mu_{eff}\simeq\sqrt{8C}$) for the Sc and the Y-compounds
were $1.4\mu_{B}$ and $0.3\mu_{B}$ which are much smaller than that
corresponding to (say) $S=1/2$. In other iridate systems as well,
effective paramagnetic moments much smaller than the pure spin value
have been observed \cite{Dey-PRB-2012-Ba3IrTi2O9,Cao-PRB-57-1998}.
This is believed to happen due to partial cancellation of the spin
and orbital part coupled via SOC \cite{Chen-PRB-82-2010}. The $\theta_{\mathrm{CW}}$
values obtained from the fit are $-588$\,K and $\sim0$K for the
Sc and Y-compounds, respectively.
\item The measured heat capacity $C_{\mathrm{P}}(T)$ (which includes the
lattice contribution) shows a sharp anomaly at about $4$K for the
Y-compound but only a weak broad hump around $10$K for the Sc-compound
(Fig. \ref{fig:CombinedHC}(a)). This suggests a transition to long-range
order in the former compound. A good fit to the high-$T$ data was
obtained using a combination of one Debye term and 3 Einstein terms
with coefficients in the ratio 1:1:5:8. The magnetic heat capacity
$C_{\mathrm{M}}(T)$ was obtained upon subtracting this lattice contribution
(Fig. \ref{fig:CombinedHC}(b)). Subsequently, integrating $C_{\mathrm{M}}(T)$
gave the entropy change ($\triangle S_{\mathrm{M}}$) associated with
the magnetic ordering for both the systems. The values of $\triangle S_{\mathrm{M}}$
obtained from our analysis are $2$J/mole K and $3.6$J/mole K for
the Sc-compound and the Y-compound, respectively. These are similar
to those obtained in Ref.\cite{Sakamoto-JSSC-179-2006,Doi-JPCM-16-2004}.
The expected value of the entropy change $R$ln($2S+1$) is $5.76$J/mol-K
for $S=1/2$, $9.13$J/mol-K for $S=1$ and $11.52$J/mol-K for $S=3/2$.
A decrease in the entropy change has also been observed in other iridate
systems such as Sr$_{2}$IrO$_{4}$ \cite{Chikara-PRB-2009-Sr2IrO4}
and Na$_{2}$IrO$_{3}$ \cite{Singh-PRB-2010-Na2IrO3}. Finally, although
the systems are very similar, the entropy change for Ba$_{3}$ScIr$_{2}$O$_{9}$
is only $\sim50\%$ of the entropy change in Ba$_{3}$YIr$_{2}$O$_{9}$.
\end{enumerate}
\begin{table}
\caption{\label{tab:Nucleus}Nuclear parameters of $^{45}$Sc and $^{89}$Y}

\centering{}%
\begin{tabular}{|c|c|c|c|}
\hline 
Nucleus & Spin ($I$) & Gyromagnetic ratio $\gamma/2\pi$  & Natural abundance\tabularnewline
\hline 
\hline 
$^{45}$Sc & $7/2$ & $10.343$\,MHz/T & $100\%$\tabularnewline
\hline 
$^{89}$Y & $1/2$ & $2.0859$\,MHz/T & $100\%$\tabularnewline
\hline 
\end{tabular}
\end{table}

\begin{figure}
\raggedright{}\includegraphics[scale=0.6]{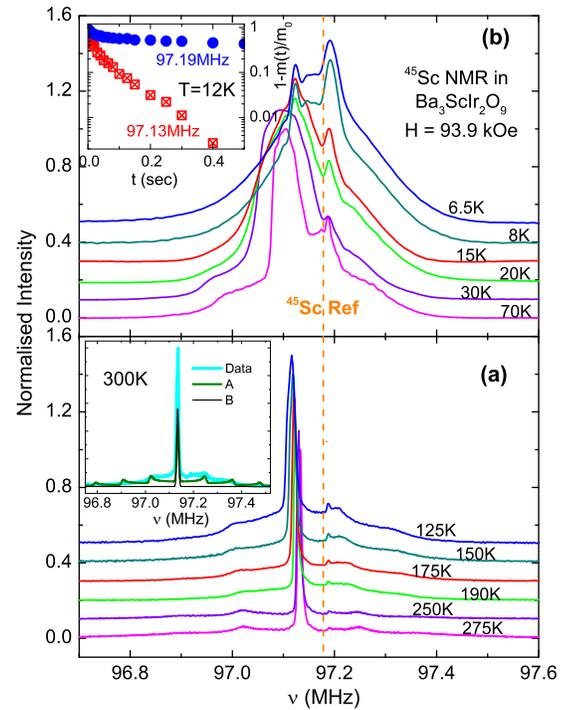}\caption{\label{fig:Ba3ScI2rO9-CombinedSpectra} The $^{45}$Sc NMR spectra
of Ba$_{3}$ScIr$_{2}$O$_{9}$ at different temperatures are shown.
The lower panel (a) shows the higher temperature ($T\geq125$K) spectra
and the upper panel (b) shows lower temperature ($T\leq70$K) ones.
The individual spectrum are shifted vertically by $0.1$ from the
previous one to have clarity. The dashed lines indicate the $^{45}$Sc
reference frequency. Lower inset: Spectrum at $300$K is shown as
thick blue line. The spectrum could be simulated by a combination
of a quadrupole split powder pattern (A) and a single line (B). The
amplitude ratio of A and B is $0.45:0.55$ (see text). Upper inset:
Longitudinal nuclear magnetization recovery curves measured at different
transmitter frequencies at $12$K.}
\end{figure}

\section{NMR results and discussion}

After some basic characterization, we have performed $^{45}$Sc NMR
and $^{89}$Y NMR studies on Ba$_{3}$ScIr$_{2}$O$_{9}$ and Ba$_{3}$YIr$_{2}$O$_{9}$
in the \emph{T} range $4-300$K which is discussed below. The different
parameters for $^{45}$Sc and $^{89}$Y nuclei are shown in Table
\ref{tab:Nucleus}.

\subsection{$^{45}$Sc NMR in Ba$_{3}$ScIr$_{2}$O$_{9}$}

First, let us consider the spectrum at $300$K. Since $^{45}$Sc has
nuclear spin $I=7/2$, one expects the spectrum to be a quadrupolar
powder pattern. However, the spectrum at room temperature (inset of
Fig. \ref{fig:Ba3ScI2rO9-CombinedSpectra}(a)) has a symmetric central
line (\textcolor{black}{shifted negatively, i.e., towards lower frequencies,
with respect to the reference frequency}) with full width at half
maxima (FWHM) about $10$kHz while the satellite peaks have a much
reduced intensity. It appears that the electric field gradient is
not very significant as $^{45}$Sc is in an octahedral environment
which is not strongly distorted. The observed spectrum at $300$K
is compared with a combination of a quadrupole split powder pattern
(spectrum A) with $\nu_{Q}=230$kHz and another one with only a coincident
central line (spectrum B) with $\nu_{Q}=0$ (see inset of Fig. \ref{fig:Ba3ScI2rO9-CombinedSpectra}(a)).
Similar analysis is also done for $^{45}$Sc NMR spectra in\textbf{
}Ba$_{3}$Cu$_{3}$Sc$_{4}$O$_{12}$ \cite{Koti-JPCM-Ba3Cu3Sc4O12}.
The mismatch in the intensity in the frequency region between the
central line and the satellite anomaly might imply that there is a
distribution of environments for $^{45}$Sc nuclei, ranging from a
nearly perfect octahedral symmetry to those with lattice distortions. 

\begin{figure}
\centering{}\includegraphics[scale=0.3]{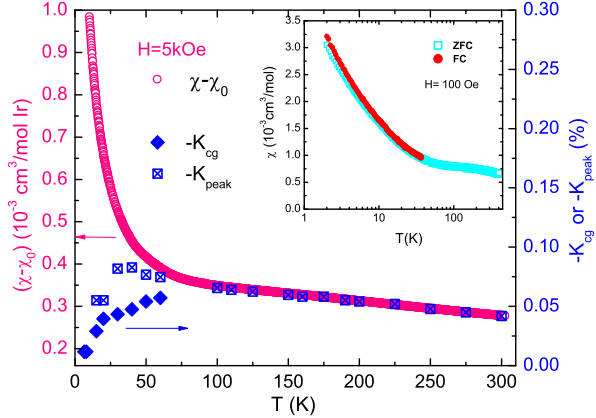}\caption{\label{fig:Ba3ScIr2O9-Kchi}The bulk susceptibility ($\chi-\chi_{0})$
of Ba$_{3}$ScIr$_{2}$O$_{9}$ is shown as function of temperature
on the left axis. The temperature variation of $^{45}$Sc NMR shift
obtained from center-of-gravity ($-K_{cg}$) and also from the peak
position ($-K_{peak}$) (at low temperature) are shown on the right
axis. Inset: The ZFC and FC susceptibilities of Ba$_{3}$ScIr$_{2}$O$_{9}$
are shown as a function of temperature in a semi-log scale.}
\end{figure}

We next look at the the evolution of the spectrum as a function of
temperature which is shown in Fig. \ref{fig:Ba3ScI2rO9-CombinedSpectra}.
As is clear, with decreasing temperature, the peak moves to lower
frequencies and the spectrum broadens. Since the peak shifts negatively
as temperature is decreased while $\chi(T)$ increases in a CW manner,
it appears that the position of the central line actually tracks $\chi(T)$
with a negative hyperfine coupling. This is illustrated in Fig. \ref{fig:Ba3ScIr2O9-Kchi}
where we have plotted the bulk susceptibility ($\chi-\chi_{0})$ as
a function of temperature on the left axis while $-K_{cg}$ or $-K_{peak}$
are plotted on the right axis. Here the shift is expressed in percent
as, for instance, $K_{peak}=\frac{\nu_{peak}-\nu_{0}}{\nu_{0}}\times100$
where $\nu_{peak}$ is the peak frequency and $\nu_{0}$ is the reference
frequency. For temperatures above $100$K, the shift $K_{peak}$ is
obtained from the maximum of the central line. At lower temperatures,
another peak begins to develop near the reference frequency and so
we have plotted both $-K_{peak}$ and $-K_{cg}$, the latter obtained
from the center of gravity of the spectrum. It is seen that both $-K_{peak}$
and $-K_{cg}$ decrease towards zero at low temperature, in contrast
to the bulk susceptibility. The upturn seen in $\chi(T)$ must then
be of extrinsic origin. By subtracting the instrinsic susceptibility
(inferred from the NMR shift) from the measured one, we obtain the
Curie behaviour due to extrinsic paramagnetic impurities which are
estimated to be about 3\% of $S=1/2$ entities. Note that the peak
in the $^{45}$Sc spectra near the reference frequency can not be
from some impurity phase since its relative intensity increases at
low temperature\textcolor{black}{{} \cite{footnote}.} These observations
suggest that the system progressively goes to a nonmagnetic state
at low temperatures. From the plot of $-K_{cg}$ as a function of
($\chi-\chi_{0})$ (not shown), which is linear in the temperature
range $75-300$K, we have calculated $A_{hf}=-2.994\mathrm{kOe}/\mu_{\mathrm{B}}$.
Finally, note that there is an abrupt change in the lineshape around
10K, the temperature at which there is a heat capacity anomaly. Whereas
the strange lineshape below 10K could be due to some peculiar spin
order, sharp peaks at around 97.13 MHz and 97.19 MHz emerge already
above 20K and do not shift with temperature. The root cause of this
strange lineshape is not unambiguously clear at the moment. 

\begin{figure}
\centering{}\includegraphics[scale=0.3]{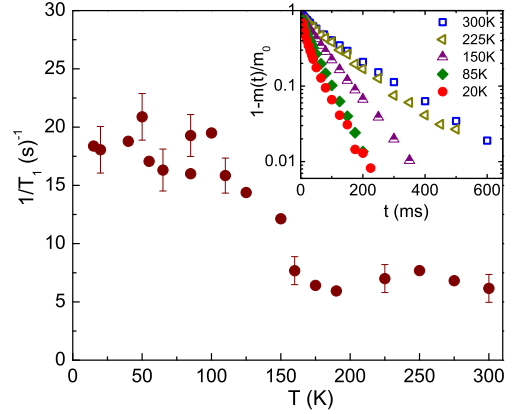}\caption{\label{fig:Ba3ScIr2O9-T1inv}The $^{45}$Sc spin lattice relaxation
rates ($1/T_{1}$) of Ba$_{3}$ScIr$_{2}$O$_{9}$ are shown as a
function of temperature. Inset: The longitudinal nuclear magnetization
recovery data for Ba$_{3}$ScIr$_{2}$O$_{9}$ are shown at a few
temperatures in a semi-log scale. The transmitter frequency was fixed
at that of the main peak, around $97.13$MHz.}
\end{figure}

Since it is clear from the discussion so far that the $^{45}$Sc nuclei
track the magnetic susceptibilities of the iridium ions, magnetic
fluctuations associated with Ir should impact the $^{45}$Sc NMR spin-lattice
relaxation rate ($1/T_{1}$) and its variation with temperature. Most
of our $1/T_{1}$ data were taken with the transmitter frequency coincident
with that of the main peak, i.e., around $97.13$MHz. A few representative
longitudinal nuclear magnetization recovery data are shown in the
inset of Fig. \ref{fig:Ba3ScIr2O9-T1inv}. All the recovery curves
have an initial fast component followed by a single exponential decay
at higher delays. Since $^{45}$Sc is a quadrupolar nucleus, in case
the central line and the satellites are not fully irradiated, the
recovery of the longitudinal nuclear magnetization is not expected
to be single exponential \cite{footnote2,Suter-JPCM-1998}. In the
present case, due to the relatively small $\nu_{Q}$, we are in a
situation of nearly full saturation of the satellites. In any case,
by measuring the slope of the long time recovery, we can extract $1/T_{1}$
which is what we have done. Temperature variation of $1/T_{1}$ is
shown in Fig. \ref{fig:Ba3ScIr2O9-T1inv}. It can be seen that from
room temperature down to $\sim175$K, $1/T_{1}$ is almost constant
as would be the case in the paramagnetic regime. Below $175$K it
increases with decreasing temperature but again becomes almost constant
below about $100$K. The increase suggests the building up of magnetic
correlations but there is an inability to order due possibly to the
geometric frustration. The leveling off near $100$K is again commensurate
with the growth of non-magnetic regions as seen from the spectra.
Also, a representative $T_{1}$ measurement taken near the reference
frequency ($97.19$MHz) indicates that the corresponding relaxation
rate is much smaller than that at the main peak (see inset of Fig.
\ref{fig:Ba3ScI2rO9-CombinedSpectra}(b)). This is further indication
that the peak near the reference frequency arises from nonmagnetic
regions. The absence of any divergence of $1/T_{1}$ with temperature
down to $4$K means that a large part of the sample remains disordered
in this temperature range. These data are somewhat reminiscent of
the LiZn$_{2}$Mo$_{3}$O$_{8}$ system where the data (a broad crossover
in the susceptibility, a hump in the specific heat, and no ordered
moments \cite{Sheckelton-NatMat-2012-LiZn2Mo3O8}) suggest a gradual
``gapping out'' rather than a phase transition. There, a physical
distortion of the triangular lattice giving rise to a decoupling of
the honeycomb lattice from the central spins has been suggested \cite{Flint-PRL-2013-LiZn2Mo3O8,Mourigal-PRL-2014-Li2ZnMo3O8}.
Whether a similar outcome (gradual gapping out in some regions of
the sample) is evident in our Sc-system is not clear though if that
were to be the case, we do not seem to sense the Curie contribution
of the central spins in the evolution of the $^{45}$Sc NMR lineshape
with temperature. 

With the above background, we  next focus on the isostructural compound
Ba$_{3}$YIr$_{2}$O$_{9}$. In this system, the aspects that are
different from the Sc-system and might be to our advantage are (i)
$^{89}$Y has nuclear spin $I=1/2$ hence there will be no quadrupolar
effect and (ii) the near-absence of a site disorder between Y and
Ir.

\begin{figure}
\centering{}\includegraphics[scale=0.35]{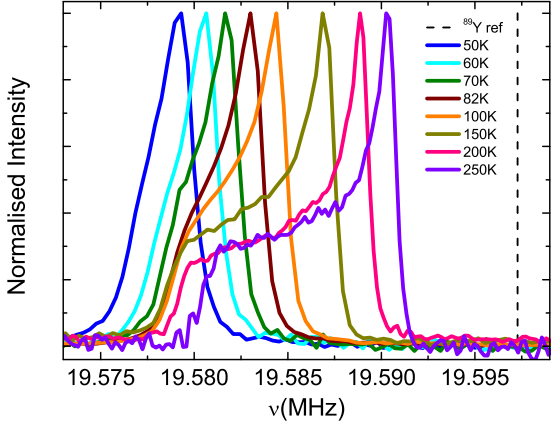}\caption{\label{fig:Ba3YIr2O9-SpectraHighT}The $^{89}$Y NMR spectral lines
in the temperature range $50-250$K. The $^{89}$Y reference position
is shown as dashed line.}
\end{figure}

\subsection{$^{89}$Y NMR in Ba$_{3}$YIr$_{2}$O$_{9}$}

Evolution of the $^{89}$Y spectral shape, shift and relaxation rates
in Ba$_{3}$YIr$_{2}$O$_{9}$ as a function of temperature are discussed
in the following. 

\begin{figure}
\centering{}\includegraphics[scale=0.35]{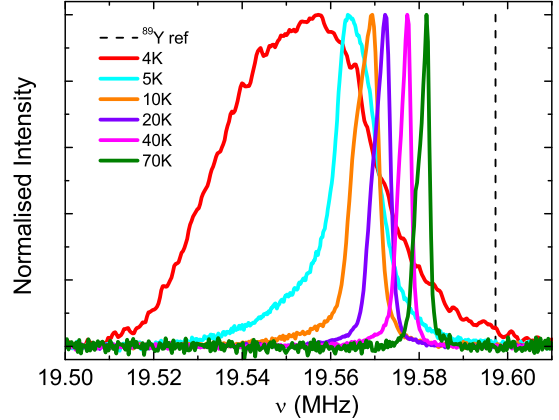}\caption{\label{fig:Ba3YIr2O9-SpectraLowT} The temperature evolution of $^{89}$Y
NMR spectra in the range $4-70$K for Ba$_{3}$YIr$_{2}$O$_{9}$
is shown. The sudden change in the spectrum at $4$K is probably due
to long range ordering. The $^{89}$Y reference is also shown as a
dashed line.}
\end{figure}

\begin{figure}
\centering{}\includegraphics[scale=0.3]{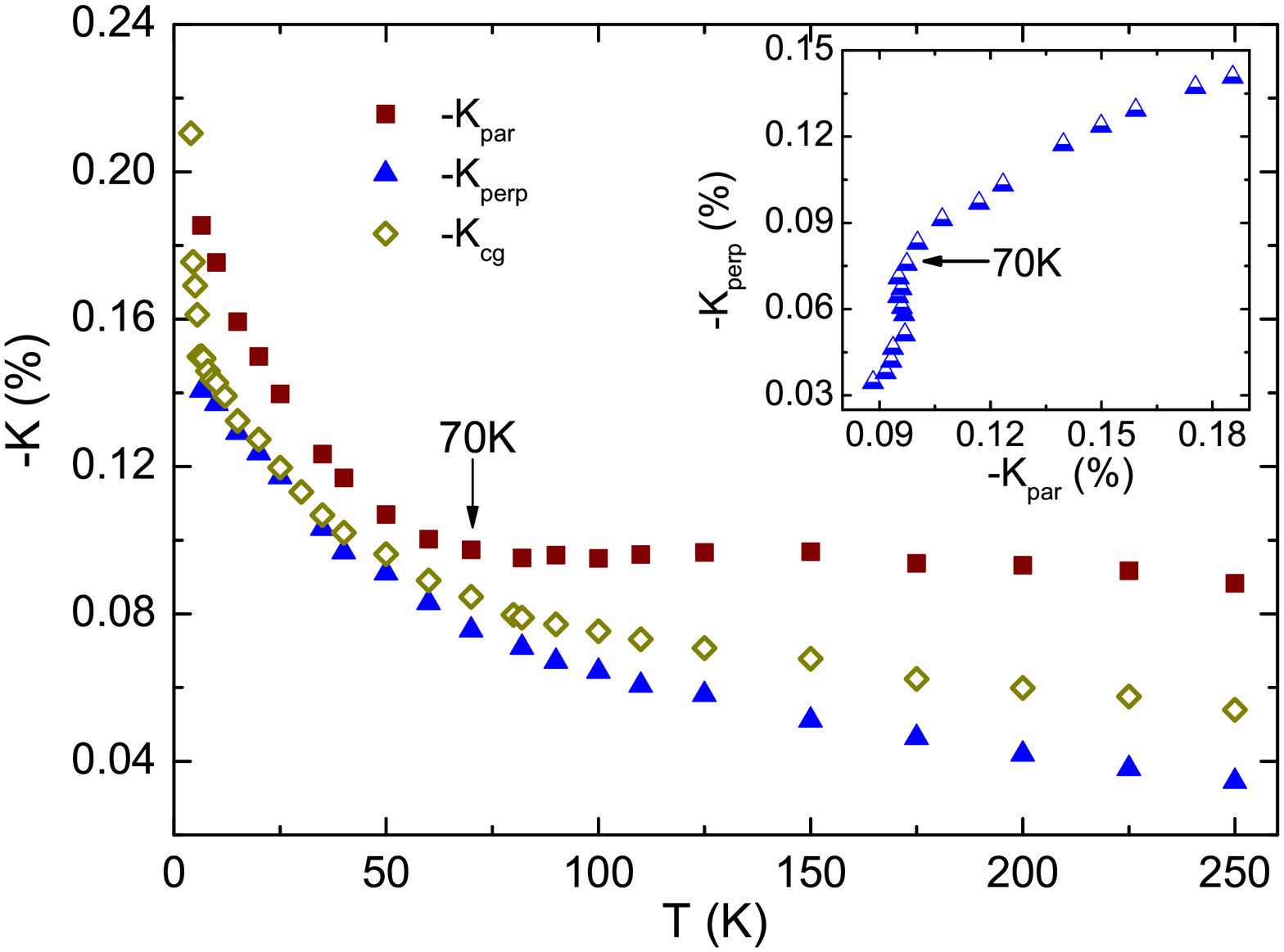}\caption{\label{fig:Ba3YIr2O9-K-anisotropy} The temperature evolution of different
shift parameters for Ba$_{3}$YIr$_{2}$O$_{9}$. Inset: $-K_{perp}$
is shown as a function of $-K_{par}$.}
\end{figure}

The $^{89}$Y spectra in Ba$_{3}$YIr$_{2}$O$_{9}$ are narrow ($\sim8-10$kHz)
except for the one at $4$K. As shown in Fig. \ref{fig:Ba3YIr2O9-SpectraHighT},
the spectral line at $250$K is asymmetric with a sharp peak on the
right side and a shoulder on the left. The spectrum at $250$K is
on the left side of the reference line and therefore has a negative
shift. Such a lineshape looks typical of a powder pattern with an
axially symmetric Knight shift tensor. However, the temperature dependence
of the anisotropic components $-K_{par}$ (contribution from $\theta=0$,
i.e., field along the axis) and $-K_{perp}$ (contribution from $\theta=90{}^{\circ}$,
i.e., field perpendicular to the axis), as shown in Fig. \ref{fig:Ba3YIr2O9-K-anisotropy},
is unusual. Although the average shift \textbf{increases} in a Curie-like
manner tracking the bulk susceptibility, $-K_{par}$ is nearly unchanged
at high-temperature while it tracks $-K_{perp}$ below $70$K or so.
This is also illustrated in $-K_{perp}$ vs. $-K_{par}$ plot (inset
of Fig. \ref{fig:Ba3YIr2O9-K-anisotropy}) which is not linear. This
would imply that either the susceptibility anisotropy or the hyperfine
coupling anisotropy changes with temperature. Another possibility
is that the anisotropy of the lineshape arises from a distribution
of magnetic environments for the $^{89}$Y nuclei. The observed variation
with temperature would then further imply that the local environments
become more homogeneous with a decrease in temperature. One could
then speculate that the distribution of magnetic environments comes
from a distortion of the triangular lattice by rotations and tilts
of the IrO$_{6}$ octahedra. One could then further speculate that
distortions heal at lower temperatures and the lattice becomes more
perfectly triangular and the system can then order (seen from the
sudden increase in linewidth below about $4$K). It is anyhow unusual
for the anisotropy of the lineshape to decrease with temperature.
In any case, this is in complete contrast to Ba$_{3}$ScIr$_{2}$O$_{9}$
where a symmetric lineshape at room temperature evolves to an asymmetric
one at lower temperatures. It should be mentioned that in the case
of Ba$_{2}$YMoO$_{6}$ also $^{89}$Y spectral lineshape was observed
to be asymmetric at high temperatures although there is a single Y
site \cite{Aharen-PRB-81-2010-Ba2YMoO6}. 

\begin{figure}
\centering{}\includegraphics[scale=0.3]{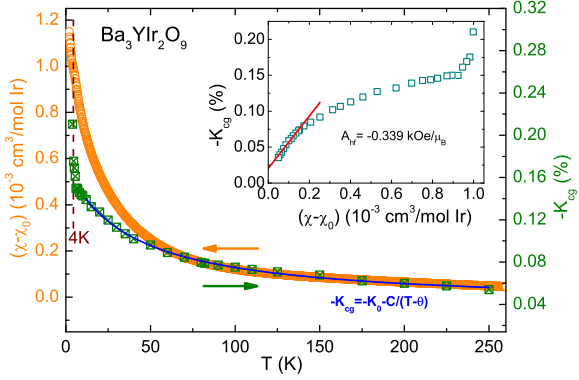}\caption{\label{fig:Ba3YIr2O9-Shift}The bulk susceptibility ($\chi-\chi_{0})$
of Ba$_{3}$YIr$_{2}$O$_{9}$ (open circles) is shown as function
of temperature on the left axis. The shift $-K_{cg}$ (green open
squares) and its fit with the Curie-Weiss formula (blue solid line)
are shown on the right axis as a function of $T$. The ordering temperature
$4$K is marked with a dashed line. Inset: $-K_{cg}$ is plotted against
($\chi-\chi_{0}$) with temperature as the implicit parameter. The
red solid line is a linear fit.}
\end{figure}

Since the spectral line is asymmetric at higher temperatures, we have
calculated the shift taking the center-of-gravity ($K_{cg}$) of each
spectrum. The temperature dependence of the shift ($-K_{cg}$) is
shown in Fig. \ref{fig:Ba3YIr2O9-Shift}. The jump in the shift near
$4$K is most likely due to the magnetic ordering. We have fitted
our shift data with the Curie-Weiss formula $-K_{cg}=-K_{0}-C/(T-\theta)$
in the temperature range $10-250$K. It is nicely fitted with the
formula and we obtained $-K_{0}=0.0387\%$ and $\theta\sim-37$K,
which means the interaction is antiferromagnetic in nature. With an
ordering temperature of $4$K, we get a frustration ratio $f=37/4>9$
where the frustration comes from the edge-shared triangular structure. 

\textcolor{black}{Next we have estimated the hyperfine coupling constant
as follows. In the inset of Fig. \ref{fig:Ba3YIr2O9-Shift}, shift
($-K_{cg}$) is plotted as a function of ($\chi-\chi_{0}$) which
is linear above $\sim60$K. From the slope of the line, we get the
hyperfine coupling $A_{\mathrm{hf}}=-0.339\mathrm{kOe}/\mu_{\mathrm{B}}$.
The actual slope has been divided by $6$ as $^{89}$Y is hyperfine
coupled to $6$ Ir. The $y-$intercept of the line gives $-K_{0}=-K_{chem}=0.021\%=210$ppm.
It should be noted that for the cubic phase sample \cite{Dey-PRB-2013-Ba3YIr}
of Ba$_{3}$YIr$_{2}$O$_{9}$, the shift is temperature independent
and has $-K_{0}=-K_{chem}=0.064\%=640$ppm. The deviation from linearity
of the $K-\chi$ data (below about $60$K) suggests a low temperature
extrinsic Curie term in $\chi(T)$. }

\begin{figure}
\centering{}\includegraphics[scale=0.3]{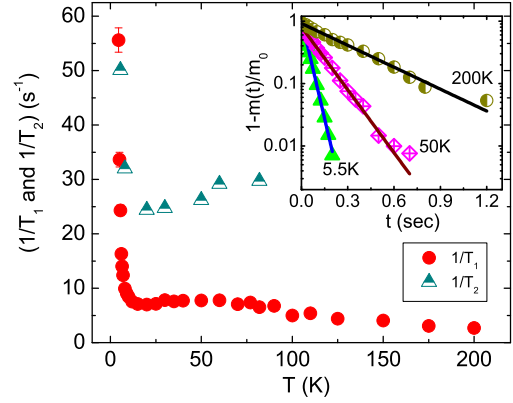}\caption{\label{fig:Ba3YIr2O9-T1T2inv}The spin-lattice relaxation rate ($1/T_{1}$)
and spin-spin relaxation rates ($1/T_{2}$) are shown as a function
of temperature. Inset: The longitudinal nuclear magnetization recovery
curves at a few temperatures are shown along with their fits (solid
line). }
\end{figure}

In contrast to the Sc-sample, the linewidth is small here and we find
a single exponential behavior for the recovery of the longitudinal
nuclear magnetization at all temperatures. A few representative recovery
curves are shown in the inset of Fig. \ref{fig:Ba3YIr2O9-T1T2inv}
with their fits. The spin lattice relaxation rate ($1/T_{1}$) is
shown as a function of temperature in Fig. \ref{fig:Ba3YIr2O9-T1T2inv}.
At $200$K, $1/T_{1}$ is small but slowly increases as temperature
is decreased. This suggests building up of short range interactions
between the spins. Once again, there is a change in the $1/T_{1}$
behavior around $70$K precisely where the lineshape becomes symmetric
and we have suggested that this reflects the ``healing'' of distortions
at low temperatures. Below about $70$K, $1/T_{1}$ becomes almost
temperature independent down to $15$K and increases below that. The
divergence in $1/T_{1}$ at $4$K is due to the critical slowing down
of the spin fluctuations on approach to long-range ordering. The temperature
variation of spin-spin relaxation rate ($1/T_{2}$) is plotted in
Fig. \ref{fig:Ba3YIr2O9-T1T2inv} along with $1/T_{1}$. At higher
temperatures $1/T_{2}$ is much larger (as might be expected) than
$1/T_{1}$ but near $4$K they become nearly equal which is another
signature of long-range ordering.

\section{conclusions}

Both the systems studied here Ba$_{3}$\emph{M}Ir$_{2}$O$_{9}$ (\emph{M}=Sc,Y)
show interesting properties. One might have expected them to be metallic
because of the extended nature of the Ir $5d$ orbitals and the fractional
valence of Ir. But in reality the samples are insulating and at least
one of them (Ba$_{3}$YIr$_{2}$O$_{9}$) orders magnetically. In
the case of Ba$_{3}$ScIr$_{2}$O$_{9}$, our local probe (NMR) data
provide evidence that the sample is magnetically homogeneous at high
temperatures, then the $^{45}$Sc nuclei have distinct magnetic environments
at intermediate temperatures and finally a large fraction of the system
goes to a singlet state at low temperatures. The associated entropy
change for Ba$_{3}$ScIr$_{2}$O$_{9}$ is almost $50\%$ of that
found in Ba$_{3}$YIr$_{2}$O$_{9}$ and much smaller than the theoretically
expected value. The data on the Sc-system suggest a gradual ``gapping
out'' as temperature is decreased though there is evidence of an
abrupt change as well in the NMR lineshape around 10K. Whether there
is any structural-distortion-assisted break-up of the triangular lattice
into singlet entities is something that remains to be proven. From
our NMR results we can say that the system is certainly interesting
for further study, say, using muon spin rotation ($\mu$SR) as a local
probe of magnetism. On the other hand, in the case of Ba$_{3}$YIr$_{2}$O$_{9}$
the $^{89}$Y lineshape is asymmetric at room temperature and progressively
becomes more symmetric at lower temperatures. This could be due to
a change, with temperature, of the anisotropy of the hyperfine couplings
or due to a change in the distribution of magnetic environments. We
feel that it is likely related to some subtle change in structural
details (such as rotation/tilt of IrO$_{6}$ octahedra) with temperature
which impacts the magnetism. The system orders magnetically at $4$K.
Nevertheless, a small Curie constant $C=0.0125$\,cm$^{3}$K/mol-Ir
($30$ times smaller than the $S=1/2$ value) and a small entropy
change is found in our measurements. These unusual features seem to
be driven by the frustration due to the triangular network of Ir-Ir
dimers in addition to a strong SOC present in the iridates.

\section{acknowledgements}

We acknowledge the financial support from Indo-Swiss Joint Research
Program, Department of Science and Technology, India. We thank H.
M. Ronnow and I. Dasgupta for useful discussions.


\begin{thebibliography}{References}
\bibitem{G.Cao book}For a review, see Frontiers of 4d- and 5d- transition
metal oxides, edited by G. Cao and L. DeLong, World Scientific (2013)

\bibitem{Kim-PRL-101-2008}B. J. Kim, H. Jin, S. J. Moon, J.-Y. Kim,
B.-G. Park, C. S. Leem, Jaejun Yu, T.W. Noh, C. Kim, S.-J. Oh, J.-H.
Park, V. Durairaj, G. Cao, and E. Rotenberg, Phys. Rev. Lett. \textbf{101},
076402 (2008)

\bibitem{Kim-Science-2009}B. J. Kim, H. Ohsumi, T. Komesu, S. Sakai,
T. Morita, H. Takagi, and T. Arima, Science \textbf{323}, 1329 (2009)

\bibitem{Cao-Na2IrO3-PRB(R)}G. Cao, T. F. Qi, L. Li, J. Terzic, V.
S. Cao, S. J. Yuan, M. Tovar, G. Murthy, and R. K. Kaul, Phys. Rev.
B \textbf{88}, 220414(R) (2013)

\bibitem{Gretarsson-PRL-2013-A2IrO3}H. Gretarsson, J. P. Clancy,
X. Liu, J. P. Hill, E. Bozin, Y. Singh, S. Manni, P. Gegenwart, J.
Kim, A. H. Said, D. Casa, T. Gog, M. H. Upton, H.-S. Kim, J. Yu, V.
M. Katukuri, L. Hozoi, J. van den Brink, and Y.-J. Kim, Phys. Rev.
Lett. \textbf{110}, 076402 (2013)

\bibitem{Manni-A2IrO3-arxiv}S. Manni, S. Choi, I. I. Mazin, R. Coldea,
M. Altmeyer, H.O. Jeschke, R. Valenti, P. Gegenwart, arXiv:1312.0815v1
(unpublished)

\bibitem{Okamoto-PRL-99-2007-Na4Ir3O8}Y. Okamoto, M. Nohara, H. Aruga-Katori,
and H. Takagi, Phys. Rev. Lett. \textbf{99}, 137207 (2007)

\bibitem{Singh-PRB-Na4Ir3O8}Y.Singh, Y. Tokiwa, J. Dong, P. Gegenwart,
Phys. Rev. B \textbf{88}, 220413(R) (2013)

\bibitem{Dey-PRB-2012-Ba3IrTi2O9}T. Dey, A. V. Mahajan, P. Khuntia,
M. Baenitz, B. Koteswararao, and F. C. Chou, Phys. Rev. B (R) \textbf{86},
140405 (2012)

\bibitem{Yang-PRB-2011-WehylMetal}K.-Y. Yang, Y.-M. Lu, and Y. Ran,
Phys. Rev. B \textbf{84}, 075129 (2011)

\bibitem{Khaliullin-PRL-d4}G. Khaliullin, Phys. Rev. Lett. \textbf{111},
197201 (2013)

\bibitem{Chaloupka-PRL-2010}J. Chaloupka, G. Jackeli, and G. Khaliullin,
Phys. Rev. Lett. \textbf{105}, 027204 (2010)

\bibitem{Wang-PRL-2011}F. Wang and T. Senthil, Phys. Rev. Lett. \textbf{106},
136402 (2011)

\bibitem{Chen-PRB-82-2010}G. Chen, R. Pereira, and L. Balents, Phys.
Rev. B \textbf{82}, 174440 (2010)

\bibitem{Kim-JSSC-177-2004-Ba3LiIr2O9}S.-J. Kim, M. D. Smith, J.
Darriet, and H.-C. zur Loye, J. Solid State Chem. \textbf{177}, 1493
(2004)

\bibitem{Sakamoto-JSSC-179-2006}T. Sakamoto, Y. Doi, and Y. Hinatsu,
J. Solid State Chem. \textbf{179}, 2595 (2006)

\bibitem{Doi-JPCM-16-2004}Y. Doi and Y. Hinatsu, J. Phys.: Condens.
Matter \textbf{16}, 2849 (2004)

\bibitem{Dey-PRB-2013-Ba3YIr}T. Dey, A.V. Mahajan, R. Kumar, B. Koteswararao,
F. C. Chou, A. A. Omrani, and H. M. Ronnow, Phys. Rev. B \textbf{88},
134425 (2013)

\bibitem{Bobroff-PRL-2009-BiCu2PO6}J. Bobroff, N. Laflorencie, L.
K. Alexander, A. V. Mahajan, B. Koteswararao, and P. Mendels, Phys.
Rev. Lett. \textbf{103}, 047201 (2009)

\bibitem{Iwase-JPSJ-Cav2O5}H. Iwase, M. Isobe, Y. Ueda and H. Yasuoka,
J. Phys. Soc. Jpn. \textbf{65}, 2397 (1996) 

\bibitem{Julien-PRL-spingap}M.-H. Julien, P. Carretta, M. Horvati\'{c},
C. Berthier, Y. Berthier, P. Ségransan, A. Carrington, and D. Colson,
Phys. Rev. Lett. \textbf{76}, 4238 (1996) 

\bibitem{Lue-PRB-2009-BiCu2VO6}C. S. Lue, S. C. Chen, C. N. Kuo,
and F. C. Chou, Phys. Rev. B \textbf{80}, 092407 (2009)

\bibitem{Libu-PRB-LiCrO2}L. K. Alexander, N. Büttgen, R. Nath, A.
V. Mahajan, and A. Loidl, Phys. Rev. B \textbf{76}, 064429 (2007)

\bibitem{Gippius-PRB-2004-LiCu2O2}A. A. Gippius, E. N. Morozova,
A. S. Moskvin, A. V. Zalessky, A. A. Bush, M. Baenitz, H. Rosner,
and S.-L. Drechsler, Phys. Rev. B \textbf{70}, 020406(R) (2004)

\bibitem{Ohsugi-PRL-1999-ordering-spinladder}S. Ohsugi, K. Magishi,
S. Matsumoto, Y. Kitaoka, T. Nagata, and J. Akimitsu, Phys. Rev. Lett.
\textbf{82}, 4715 (1999) 

\bibitem{Nath-PRB-2005}R. Nath, A. V. Mahajan, N. Büttgen, C. Kegler,
A. Loidl, and J. Bobroff, Phys. Rev. B \textbf{71}, 174436 (2005)

\bibitem{Bobroff-PRL-NMR-cuprate}J. Bobroff, H. Alloul, S. Ouazi,
P. Mendels, A. Mahajan, N. Blanchard, G. Collin, V. Guillen, and J.-F.
Marucco, Phys. Rev. Lett. \textbf{89}, 157002 (2002)

\bibitem{Nakai-PRL-2010-BaFe2As-P}Y. Nakai, T. Iye, S. Kitagawa,
K. Ishida, H. Ikeda, S. Kasahara, H. Shishido, T. Shibauchi, Y. Matsuda,
and T. Terashima, Phys. Rev. Lett. \textbf{105}, 107003 (2010)

\bibitem{Ishida-Review-FeAs}K. Ishida, Y. Nakai, and H. Hosono, J.
Phys. Soc. Japan 78, 062001 (2009) and references therein

\bibitem{Takagi-PPT}See http://online.itp.ucsb.edu/online/motterials07/takagi

\bibitem{Aharen-PRB-81-2010-Ba2YMoO6}T. Aharen, J. E. Greedan, C.
A. Bridges, A. A. Aczel, J. Rodriguez, G. MacDougall, G. M. Luke,
T. Imai, V. K. Michaelis, S. Kroeker, H. Zhou, C. R. Wiebe, and L.
M. D. Cranswick, Phys. Rev. B \textbf{81}, 224409 (2010)

\bibitem{Aharen-PRB-2009-Ba2YRuO6}T. Aharen, J. E. Greedan, F. Ning,
T. Imai, V. Michaelis, S. Kroeker, H. Zhou, C. R. Wiebe, and L. M.D.
Cranswick, Phys. Rev. B \textbf{80}, 134423 (2009)

\bibitem{Quilliam-PRL-2012-Ba3CuSb2O9}J. A. Quilliam, F. Bert, E.
Kermarrec, C. Payen, C. Guillot-Deudon, P. Bonville, C. Baines, H.
Luetkens, and P. Mendels, Phys. Rev. Lett. \textbf{109}, 117203 (2012)

\bibitem{Imai-PRL-2008}T. Imai, E. A. Nytko, B. M. Bartlett, M. P.
Shores, and D. G. Nocera, Phys. Rev. Lett. \textbf{100}, 077203 (2008)

\bibitem{Cao-PRB-57-1998}G. Cao, J. Bolivar, S. McCall, J. E. Crow,
and R. P. Guertin, Phys. Rev. B \textbf{57}, R11039 (1998)

\bibitem{Chikara-PRB-2009-Sr2IrO4}S. Chikara, O. Korneta, W. P. Crummett,
L. E. DeLong, P. Schlottmann, and G. Cao, Phys. Rev. B \textbf{80},
140407(R) (2009) 

\bibitem{Singh-PRB-2010-Na2IrO3}Y. Singh and P. Gegenwart, Phys.
Rev. B \textbf{82}, 064412 (2010)

\bibitem{Koti-JPCM-Ba3Cu3Sc4O12}B. Koteswararao, A.V. Mahajan, F.
Bert, P. Mendels, J. Chakraborty, V. Singh, I. Dasgupta, S. Rayaprol,
V. Siruguri, A. Hoser and S. D. Kaushik, J. Phys.: Condens. Matt.
\textbf{24}, 236001 (2012)

\bibitem{footnote} It is true that due to a possible contrast in
the spin-spin relaxation times $T_{2}$ for the two peaks at low temperature
(a much longer $T_{2}$ for the unshifted peak compared to the main
peak), one can artificially enhance the intensity of the peak near
the reference position. We have therefore taken care of this and kept
the time delay between the echo producing \textit{\textcolor{black}{rf}}
pulses to be several times shorter than the shortest $T_{2}$. Further,
the repetition time of the pulse sequence was several times the longest
$T_{1}$. The Sc/Ir site mixing might also contribute to the unshifted
peak though the intensity should not change with temperature. We also
confirmed that the total integrated spectral intensity times the temperature
is nearly constant below, say, 50K.

\bibitem{Flint-PRL-2013-LiZn2Mo3O8}R. Flint and P. A. Lee, Phys.
Rev. Lett. \textbf{111}, 217201 (2013)

\bibitem{Mourigal-PRL-2014-Li2ZnMo3O8}M. Mourigal, W. T. Fuhrman,
J. P. Sheckelton, A. Wartelle, J. A. Rodriguez-Rivera, D. L. Abernathy,
T. M. McQueen, and C. L. Broholm, Phys. Rev. Lett. \textbf{112}, 027202
(2014)

\bibitem{Sheckelton-NatMat-2012-LiZn2Mo3O8}J. P. Sheckelton, J. R.
Neilson, D. G. Soltan, and T. M. McQueen, Nat. Mater. \textbf{11},
493 (2012)

\bibitem{footnote2}In case, one irradiates only the central line
and the duration of the saturating sequence is much less than $T_{1}$,
the recovery law is given by 
\[
1-m(t)/m_{0}=A\,[0.011905\, exp(-t/T_{1})+0.068182\, exp(-6t/T_{1})+0.20605\, exp(-15t/T_{1})+0.713868\, exp(-28t/T_{1})]
\]
In case, the duration of the saturating comb is much greater than
$T_{1}$ (with only the central line getting saturated), then the
recovery law is 
\[
1-m(t)/m_{0}=A\,[0.190476\, exp(-t/T_{1})+0.181819\, exp(-6t/T_{1})+0.21978\, exp(-15t/T_{1})+0.40792\, exp(-28t/T_{1})]
\]
However, if the central line is fully saturated and the satellites
are also partly saturated, the coefficient of the $e{}^{-t/T_{1}}$
term will be greater than given above and will approach one for full
saturation. In case of a small weight for the $e{}^{-t/T_{1}}$ term,
recovery would have attained a linear behaviour (on a log-linear scale)
only at very long times, in contrast to our data.

\bibitem{Suter-JPCM-1998}A. Suter, M. Mali, J. Roos, and D. Brinkmann,
J. Phys.: Condens. Matter \textbf{10}, 5977 (1998)\end{thebibliography}
\end{document}